\documentstyle[12pt]{article}
\newfont{\ffont}{msym10}                        %%
\newcommand{\beq}{\begin{equation}}             %%
\newcommand{\eeq}{\end{equation}}               %%
\newcommand{\bqry}{\begin{eqnarray}}            %%
\newcommand{\eqry}{\end{eqnarray}}              %%
\newcommand{\bqryn}{\begin{eqnarray*}}          %%
\newcommand{\eqryn}{\end{eqnarray*}}            %%
\newcommand{\preprint}[1]{\begin{table}[t]      %%
            \begin{flushright}                  %%
            \begin{large}{#1}\end{large}        %%
            \end{flushright}                    %%
            \end{table}}                        %%
\newcommand{\PD}[2]                             %%
    {\frac{\partial^{#2}}{\partial #1^{#2}}}    %%
\begin{document}
\preprint{LA-UR-95-4527}
\title{First Order Quark-Gluon/Hadron Transition
 May Affect Cosmological Nucleosynthesis}
\author{\\ L. Burakovsky\thanks {Bitnet:
BURAKOV@QCD.LANL.GOV.} \ \\   \\ Theoretical Division,
T-8 \\ Los Alamos National Laboratory \\ Los Alamos,
NM 87545, USA \\}
\date{ }
\maketitle
\begin{abstract}
In the model of a first order quark-gluon/hadron phase
transition in which the hadronic phase is considered
as vacuum bubbles growing in the quark-gluon background
with chiral symmetry broken inside the bubble, we find
the estimate for the length scale associated with
inhomogeneities originated during the transition, $10$
m$\;\stackrel{<}{\sim }\ell \stackrel{<}{\sim }40$ m,
being sufficient to produce significant effects on
cosmological nucleosynthesis.
\end{abstract}

\bigskip
\section*{  }
Recent lattice gauge simulations indicate that the pure
$SU(3)$ gauge theory has a first order phase transition
between the low temperature color confining hadronic
phase and the high temperature deconfined plasma phase
\cite{Pet}. Because of the first order nature of the
phase transition, these two phases can coexist at the
critical temperature with a distinct phase interface
in between; this interface carries a positive amount of
extra free energy, i.e. a surface tension ($\sigma ).$

The surface tension is a fundamental parameter in the
description of the time evolution of hot hadronic
matter as it proceeds through the phase transition.
In the scenario of a homogeneous nucleation \cite{LL}
the probability for a bubble to nucleate is
\beq
p(T)=\exp \left\{ -\frac{16 \pi \sigma ^3}{3T_c
(p-p^{'})^2}\right\}.
\eeq
In this model, the bubble at nucleation time has a
radius large enough to be hydrostatically stable in
the supercooled plasma, and to sustain a difference
in pressure $(p-p^{'})$ between two phases.

The question has been raised in the literature
\cite{lit,FMA} whether baryon inhomogeneities could
originate during a first order quark-gluon/hadron
phase transition that would alter the successful
conventional calculations of cosmological
nucleosynthesis \cite{OSSW}. An essential ingredient
in these calculations is the surface tension for
bubbles of the new hadronic phase. The length scale
associated with the inhomogeneities has been given
in ref. \cite{FMA} \footnote{Eq. (2) takes into
account the correction given in ref. \cite{KKR}.}
(we will arrive at this relation below):
\beq
\ell \simeq (3.4\times 10^4\;{\rm m})(\sigma /
{\rm MeV}^3)^{3/2}(T_c/{\rm  MeV})^{-13/2},
\eeq
where $T_c$ is the critical temperature for a first
order quark-gluon/hadron phase transition. Campbell
{\it et al.} \cite{Cam} have obtained an estimate
for this surface tension in the framework of the
effective low energy Lagrangian for broken chiral
and scale invariance \cite{GPY} in the hadronic phase,
\beq
\sigma \simeq (70\;{\rm MeV})^3,
\eeq
which is in good agreement with the estimate made in
ref. \cite{AF}. Recently the surface tension has been
calculated in quenched QCD by lattice numerical methods
\cite{KKR,HPRS,BHPR}, as well as analytically in the
mean-field approximation \cite{FP,Tanmoy}. When
performed on $N_t=2$ lattices, the numerical studies
have produced the results \cite{KKR}
\beq
\sigma /T^3=0.24(6)
\eeq
and \cite{HPRS}
\beq
\sigma /T^3=0.12(2),
\eeq
while for $N_t=4$ lattice \cite{BHPR}
\beq
\sigma /T^3=0.027(4).
\eeq
With $\sigma $ given by (3)-(6) and $T_c\simeq 100$ K,
the formula (2) gives the estimate $\ell \simeq 0.5-1$
m, which is hardly of interest for nucleosynthesis
since significant effects on nucleosynthesis require
$\ell >10$ m and probably $\ell >100$ m.\footnote{As
shown in ref. \cite{AFMM}, $\ell \simeq 30$ m is
needed for the most interesting models of a
non-homogeneous universe.} Because of a strong
dependence of $T_c,$ $\ell $ will be smaller for
larger $T_c;$ e.g., for $T_c\simeq 150$ K, Eq. (2)
with $\sigma $ given by (3) yields $\ell \simeq 0.1$ m.

Bhattacharya {\it et al.} \cite{Tanmoy} have found the
expression for the interface tension between distinct
$Z(N)$ vacua above the deconfinement transition for a
pure $SU(N)$ gauge theory $(N=2,3),$
\beq
\sigma = \frac{4(N-1)\pi ^2}{3\sqrt{3N}}\frac{T^3}{g},
\eeq
where $g$ is the gauge coupling constant. For a typical
value of $g\simeq 1-2$ (corresponding to $\alpha _s\equiv
g^2/4\pi \simeq 0.1-0.3)$ at temperature of the order of
a typical deconfinement one, $T\simeq 150$ MeV \cite{Ber},
the formula (2) yields the estimate
\beq
\ell \simeq (15-30)\;{\rm m},
\eeq
which is considerably larger than that obtained from Eqs.
(3)-(6) above.

In this letter we show that the estimate for $\ell $
following from the values of the surface tension
obtained in refs. \cite{KKR,Cam,AF,HPRS,BHPR}, $\ell
\simeq 0.5-1$ m, should be raised by one-two order of
magnitude, in agreement with Eq. (8), if one considers
the new hadronic phase as vacuum bubbles growing in
the quark-gluon environment, with chiral symmetry
broken (i.e., nonzero value of the quark condensate)
inside the bubble.

At temperatures well below $T_c,$ hadronic matter
consists of the lightest hadrons, the pions. At $T$
near $T_c,$ the $\pi \pi $ interaction becomes strong,
the $\pi \pi $ amplitudes at the relevant energies
become so large that they allow for some bound states,
the resonances \cite{Shu}. The method for taking into
account such resonance interaction was suggested by
Belenky and Landau \cite{BL} as considering the
unstable particles on an equal footing with the
stable ones in the thermodynamic quantities, by means
of a resonance spectrum. Such a spectrum in both the
statistical bootstrap model \cite{Hag, Fra} and the
dual resonance model \cite{FV} takes on the form
\beq
\rho (m)\sim m^a\;e^{m/T_0},
\eeq
where $a$ and $T_0$ are constants. This treatment of
a hadronic resonance gas leads to a singularity in the
thermodynamic functions at $T=T_0$ \cite{Hag,Fra} and,
in particular, to an infinite number of the effective
degrees of freedom in the hadronic phase, thus making
a transition to the quark-gluon phase impossible. To
cope with this difficulty, it is normally suggested
that the hadrons are strongly interacting particles
with finite range interactions among them which
increase with increasing density. These interactions
cannot be ignored in discussing the thermodynamics of
hadronic matter; in fact, it has been shown \cite{Cley}
that the neglect of such interactions leads to
unphysical behavior for a deconfinement phase
transition mentioned above. (The mass spectrum (9)
originates from the clustering of hadrons \cite{Fra};
such a clustering corresponds to an effective
attractive interaction and would not allow for a
transition to the quark-gluon phase since for this
transition a repulsive interaction at small distances
is necessary.) In ref. \cite{Cam} the hadronic
resonance spectrum (9) was modified to include an
effective repulsive interaction by means of i) a
hard-core potential, ii) an excluded volume
(Van der Waals approach). It has been shown that in
both cases the number of the effective degrees of
freedom in the hadronic phase is now finite (namely,
6.5-7, as read off from Figs. 6,7 of ref. \cite{Cam}),
and a normal (first order) transition to the
quark-gluon phase becomes possible. For our
present purposes we shall restrict ourselves to the
pions alone (i.e., to the 3 effective degrees of
freedom), and think of the hadronic phase as that of
the pions which are brought to the boiling point at
$T_c$ with the energy density $\rho _\pi (T_c)=3
\pi ^2/90\;T_c^4.$

The usual scheme of a first order phase transition
is as follows. With the increase of energy
deposition (e.g., in relativistic heavy ion
collisions), bubbles of the quark-gluon plasma
``vapor'' are formed until at $\rho \sim g/3\;\rho _
\pi (T_c)$ the system is entirely vapor ($g$ being
the total number of degrees of freedom in the
quark-gluon phase, which is 37 for two-flavor
plasma and 47.5 for three-flavor one). The
quark-gluon to hadron transition in the early
universe proceeds in a slightly different way:
when the universe is about 10 $\mu $sec old,
its temperature drops down to $T_c\simeq 150$ MeV.
However, the phase transition does not start
immediately but the interface tension causes
supercooling by the amount of $\triangle T/T_c
\simeq 0.02(\sigma /T_c^3)^{3/2}$ \cite{KK}.
After the period of supercooling the hadron
bubbles start nucleating; this process gives
rise to shock waves which reheat the universe
back to $T_c,$ thus hindering the further
formation of the hadron bubbles \cite{KK}
(in a different scenario, bubble formation
and growth is sufficient to begin reheating
the system due to the release of latent heat
\cite{CK}). The transition continues because
of the growth of previously created bubbles,
until the system gets entirely hadronized.
The average distance between the bubbles
turns out to be \cite{FMA}
\beq
\ell \simeq 10^{-4}\;R_{H}(\sigma /T_c^3)^{
3/2}\;{\rm m},
\eeq
where $R_{H}$ is the horizon size at the
transition time,
\beq
R_{H}\sim \frac{3\cdot 10^8\;{\rm m}}{(T_c/{
\rm MeV})^2}\sim 10\;{\rm km}.
\eeq
In view of (10),(11), one obtains the estimate
(2) for the scale of inhomogeneities in the
hadron distribution which may have affected
cosmological nucleosynthesis.

The transition temperature is calculated by
equating pressures. If the quark-gluon plasma
is treated within the bag model, then
\beq
p_\pi =3\frac{\pi ^2}{90}T_c^4=p_{QGP}=g\frac{
\pi ^2}{90}T_c^4-B
\eeq
$(B$ being the bag constant, $B\simeq (245-255\;
{\rm MeV})^4$ \cite{BBP}), giving the value of
$T_c$ $\simeq 170$ MeV for two-flavor plasma and
$\simeq 160$ MeV for three-flavor one.

In conventional bag model the difference between
the quark condensate values on different sides
of a bag surface is normally disregarded. To
estimate the effect of the quark condensate
discontinuity on the properties of hadrons,
we treat the letter within the bag model and
use the energy-momentum tensor $T^{\mu \nu}_q$
of the quark fields $\psi ^a_f(x)$ with massless
quarks inside the bag. In zero order
approximation of a power expansion in the
QCD coupling constant $\alpha _s$ the
following expression holds:
\beq
T^{\mu \nu }_q = \frac{i}{2}\sum _f\left(
\bar{\psi }^a_f\gamma ^\mu \partial ^\nu
\psi ^a_f - (\partial ^\nu \bar{\psi }^a_f)
\gamma ^\mu \psi ^a_f \right),
\eeq
where $a=1,2,3$ is SU(3)$_c$ color index, and
$f=u,d,s$ is SU(3)$_f$ flavor index. The
linear boundary condition\footnote{In the
chiral bag model, the boundary condition
reads $i\gamma ^\mu n_\mu \psi ^a_f(x)=e^{i
\tau ^b\pi ^b \gamma ^5}\psi ^a_f(x),$ where
$b=1,2,3$ is the isospin index, and couples
the quark fields on one side of the bag
boundary with the pion field on the other
side in a highly nonlinear manner. We do
not discuss this model here.} \cite{Ch}
\beq
i\gamma ^\mu n_\mu \psi ^a_f(x)=\psi ^a_f(x)
\eeq
on the bag surface with external normal $n_\mu $
corresponds to the relation $\bar{\psi }^a_f\psi
^a_f=0$ on the interior side of the bag surface.
On the exterior side $\bar{\psi }^a_f\psi ^a_f
\neq 0$ because the corresponding vacuum
condensates have nonzero values \cite{SVZ}.
In this case the quark field contributions to
the bag energy density $\rho _q=T^{00}_q$
contain not only conventional part corresponding
to the quark kinetic energy but also contain the
usually ignored surface part. The surface part
arises from the contribution of the discontinuity
of the quark condensate values on both sides of
a bag surface \cite{Pok}:
\beq
E_s=-\frac{1}{4}S\sum _f\langle \bar{\psi }^a_f
\psi ^a_f\rangle =\sigma _{vac}S.
\eeq
Here $S$ is a surface area of the bag, and
\beq
\sigma _{vac}\equiv -\frac{1}{4}\sum _f\langle
\bar{\psi }^a_f\psi ^a_f \rangle.
\eeq
Typical ``empirical'' values discussed in the
literature are \cite{SVZ,RRY}
\beq
\langle \bar{\psi }^a_u\psi ^a_u\rangle \simeq
\langle \bar{\psi }^a_d\psi ^a_d\rangle \simeq
-(240 \pm 25\;{\rm MeV})^3,
\eeq
$\langle \bar{\psi }^a_s\psi ^a_s\rangle $ being
of similar magnitude. The commonly adopted value
of the quark condensate for calculations within
the framework of QCD sum rules is \cite{AK}
\beq
\langle \bar{\psi }^a\psi ^a\rangle =-0.0241\;
{\rm GeV}^3\cong -(289\;{\rm MeV})^3.
\eeq
Sometimes people consider even higher values of
the quark condensate, e.g. \cite{LKW}
$$\langle \bar{\psi }^a_u\psi ^a_u\rangle =
\langle \bar{\psi }^a_d\psi ^a_d\rangle =
-(287\;{\rm MeV})^3,$$
\beq
\langle \bar{\psi }^a_s\psi ^a_s\rangle =
-(306\;{\rm MeV})^3,
\eeq
or \cite{ZHK}
\beq
\langle \bar{\psi }^a\psi ^a\rangle \simeq -0.032
\;{\rm GeV}^3\cong -(315\;{\rm MeV})^3.
\eeq
We shall take
\beq
\langle \bar{\psi }^a_u\psi ^a_u\rangle \simeq
\langle \bar{\psi }^a_d\psi ^a_d\rangle \simeq
\langle \bar{\psi }^a_s\psi ^a_s\rangle \simeq
-(270-280\;{\rm MeV})^3,
\eeq
which lies somehow between the values provided
by (17)-(20). Then the value of $\sigma _{vac}$
calculated from Eq. (16) is
\beq
\sigma _{vac}\simeq (245-255\;{\rm MeV})^3.
\eeq
This surface tension is strong in comparison
with the vacuum pressure (the bag constant)
\footnote{The estimate that we obtained,
$\sigma ^{1/3}\approx B^{1/4},$ was also
suggested in ref. \cite{FJ}.} \cite{BBP}
\beq
B=-\rho _{vac}=\frac{9}{32}\langle \frac{
\alpha _s}{\pi }G^a_{\mu \nu }G^{a \mu \nu }
\rangle -\frac{1}{4}\sum _fm_f\langle \bar{
\psi }^a_f\psi ^a_f\rangle \simeq (245-255\;
{\rm MeV})^4.
\eeq
Here $G^a_{\mu \nu }$ is a gluon field stress
tensor and $m_f$ is a typical bare (current)
quark mass.

Now consider the hadronic phase arising as a
result of a first order quark-gluon/had- ron
transition in the form of vacuum bubbles
growing in the bulk quark-gluon plasma.
Such a vacuum bubble is a sperical surface
with quarks and gluons outside it obeying the
bag boundary conditions\footnote{The bag
boundary conditions for gluons are discussed
in ref. \cite{MS}.} (14). Inside the bubble
we have nonzero quark $\langle \bar{\psi }^
a_f(x)\psi ^a_f(x)\rangle $ and gluon
$\langle G^a_{\mu \nu }G^{a\mu \nu }\rangle $
condensates. Therefore, one can apply arguments
similar to the previous ones and obtain the
same expression for the interface surface
tension as above (Eq. (16)). Thus, the value
of $\sigma $ given by Eq. (22) should be
used as the realistic value for the bubble
surface tension  in the expression for the
length scale associated with the
inhomogeneities, Eq. (2). The exect value
for the critical temperature of a
quark-gluon/hadron phase transition is not known.
In the framework of chiral perturbation theory,
Gerber and Leutwyler \cite{GL} have calculated
the temperature of a chiral symmetry restoration
transition (to be associated with the hadron to
quark-gluon one) $T_c\simeq 170$ MeV in the
SU(2)$_f$ chiral limit. The value of the
critical temperature provided by lattice
gauge simulations is currently \cite{Ber}
$T_c\simeq 140$ MeV. We, therefore, consider
the value of the critical temperature of a
quark-gluon/hadron transition to be in the
temperature interval
\beq
140\;{\rm MeV}\stackrel{<}{\sim }T_c\stackrel{
<}{\sim}170\;{\rm MeV},
\eeq
as granted by both chiral perturbation theory
and the most recent lattice simulations (note
that both values of $T_c$ estimated above
within the bag model lie in this temperature
range). For the temperature interval (24) and
the value of the bubble tension given by (22),
Eq. (2) gives
\beq
10\;{\rm m}\stackrel{<}{\sim }\ell \stackrel{
<}{\sim }40\;{\rm m},
\eeq
which is one-two order of magnitude as much as
the estimate following from the values of
$\sigma $ found in refs.
\cite{KKR,Cam,AF,HPRS,BHPR}. In view of the
aforementioned estimate $\ell >10$ (or 100) m,
this value of the length scale should be
considered as being sufficient to produce
significant effects on nucleosynthesis.
Note that the estimates of refs.
\cite{Tanmoy,AFMM} lie in the interval (25).

By further adjustment of the values of $\sigma $
and $T_c$ the estimate (25) may be raised or
diminished, respectively. For example, for
$T_c\simeq 100$ MeV and $\sigma $ given by (22)
one obtains $\ell \simeq 200-250$ m.

The question has been raised in ref. \cite{SFT}
whether inhomogeneities originating during a
first order quark-gluon/hadron transition can
produce cosmological beryllium or boron. In
contrast to simple inhomogeneities like those
with $\ell \simeq 0.5-1$ m, inhomogeneities
associated with the length scale (25) are
likely to produce cosmological Be or B. If
this is really the case, it will have great
implications for Big Bang nucleosynthesis.
Cosmological Be or B are a signature for
significant density variations \cite{SFT}.
Such variations could lead to planetary mass
black holes \cite{CS} or quark nuggets
\cite{nug,Wit} that could serve as the dark
matter of the universe.

As remarked in \cite{Pet}, another
interesting point which may lead to a
non-standard scenario of nucleosynthesis
is that there may be a baryon number
contrast in the hadron and the quark-gluon
plasma phases. The thermodynamic advantage
to place most of the baryon number into the
quark-gluon phase was analyzed in ref.
\cite{AFMM}. A large penalty in the free
energy is paid when a unit of baryon number
is placed in the hadron phase because of the
large mass of the baryons, as compared to the
phase of massless quarks for which there is no
mass penalty in the free energy. This may not
be the case if an effective baryon mass is
temperature dependent and drops significanfty
as $T$ gets closer to $T_c,$ as suggested in a
series of papers by Brown and Rho \cite{BR}.
It is most probably that the estimate of ref.
\cite{AFMM} for the baryon number contrast in
two phases, $R\sim O(100),$ should be
diminished by one order of magnitude. Even in
this case, the baryon number contrast in two
phases will still remain large.

Two phenomena, the large inhomogeneities scale in
the hadron distribution and the large baryon number
contrast in the hadron and the quark-gluon plasma
phases should lead to sizable effects on the
standard scenario of cosmological nucleosynthesis.
With the observations made in this letter,
conventional calculations of cosmological
nucleosynthesis should be amended by nontrivial
details. New work on this and related subjects is
in progress.

\section*{Acknowledgements}
I wish to thank T. Bhattacharya, L.P. Horwitz and
E. Mottola for very valuable discussions.

\bigskip
\bigskip
%\newpage

\end{document}